# A fast low-to-high confinement mode bifurcation dynamics in a tokamak edge plasma gyrokinetic simulation

C.S. Chang[a], S. Ku[a], G.R. Tynan[b], R. Hager[a], R.M. Churchill[a], I. Cziegler[b,†], M. Greenwald[c], A. E. Hubbard[c], J. W. Hughes[c]

[a]*Princeton Plasma Physics Laboratory*
[b]*University of California San Diego, CA*
[c]*MIT Plasma Science and Fusion Center*

Transport barrier formation and its relation to sheared flows in fluids and plasmas are of fundamental interest in various natural and laboratory observations and of critical importance in achieving an economical energy production in a magnetic fusion device. Here we report the first observation of an edge transport barrier formation event in electrostatic gyrokinetic simulation carried out in a realistic diverted tokamak edge geometry under strong forcing by a high rate of heat deposition. The results show that turbulent Reynolds-stress driven sheared ExB flows act in concert with neoclassical orbit loss to quench turbulent transport and form a transport barrier just inside the last closed magnetic flux surface.

Transport barrier formation and its relation to the flow of the fluid medium are of fundamental interest in various natural and laboratory observations, such as geophysical and atmospheric fluid systems, *etc*. [1, 2]. In a magnetic fusion device, this physics has a critical implication to achieving an economical energy production since the bifurcated plasma state, called high confinement mode (H-mode) [3], is often envisioned as the operating mode of choice for fusion reactors [4], and will be relied on in ITER in achieving its goal of ten-fold energy gain [5]. However, despite over 30 years of H-mode operation, there has been no fundamental understanding at the kinetic level on how the H-mode bifurcation occurs.

Experimentally, a radial transport bifurcation into the H-mode in both plasma density and thermal channels occurs in a thin edge layer of the tokamak plasma just inside the magnetic separatrix surface when the plasma heating power exceeds a critical value [3]. As a result, a plasma density and temperature pedestal is formed on the time scale of a few ms with a steep gradient in the thin edge layer. As this pedestal forms the core plasma pressure inside the edge layer position increases, resulting in a transition of plasma operation to a high confinement H-mode from a low confinement L-mode [3]. The bifurcation event is accompanied by a sharp increase in the sheared ExB flow and significant drop in the turbulence amplitude within the thin transport barrier layer on a time-scale that is often shorter than 0.1ms if the heating power is strong (strongly driven). The edge heating needed to initiate this H-mode regime is minimized when the ion $\nabla B$-drift direction, or the $\vec{B} \times \nabla B$ direction, is toward the magnetic X-point when the plasma is operated with a single poloidal divertor [6].

There have been many attempts to use simple theoretical models on how an H-mode transition could occur. A popular "predator-prey" model [7] implies that increasing the heat flux to the edge of the plasma, thus raising the edge gradients, results in stronger turbulence (prey). The increased turbulence can then amplify the sheared poloidal flow (predator) nonlinearly through the turbulent Reynolds stress. When the flow drive is larger than the flow damping, the sheared poloidal flow can grow, nonlinearly extracting even more kinetic energy from the turbulence. As a result, the turbulence and the associated turbulent transport collapse. This suppressed turbulence state is then conjectured to be maintained through the steep-pressure driven sheared ExB flow driven by the simultaneous build-up of the H-mode pedestal.

Extended predator-prey models predict both oscillatory limit-cycle (LCO) type predator-prey transition [8] and a sharp transition [9, 10, 11] triggered by a single burst of axisymmetric sheared turbulence-driven ExB flow (known as zonal flow). Turbulent fluid simulations have shown evidence for some of this phenomenology [24-26]. Experiments have indeed reported both LCO type transition [12-17] when operating close to the H-mode power threshold, and a sharp bifurcation [18-21] within 0.1ms [21] when the power threshold is exceeded significantly.  In the fast transition, some detailed experiments report that the turbulent stress-driven shear flow first leads to a collapse of the turbulence, which is then followed by the development of the edge pedestal in a rather longer time scale, claiming that the turbulence suppression is not maintained by the simultaneous buildup of the steep pedestal and the associated ExB shearing [22,23]. Some experiments [i.e., 27] report a different evidence that the experimentally observed Reynolds work is too weak to explain the L-H bifurcation and, thus, the ExB shearing from the neoclassical orbit loss physics [28,29] is solely responsible for the bifurcation.

This body of evidence suggests that the H-mode transition could indeed be related to the sheared ExB flow, either turbulence or orbit-loss driven. However, the existing models are based upon simplified ad-hoc equations and the turbulence simulations assume specific instability mechanisms, ignore possible important kinetic effects, or are not carried out in a realistic geometry.

This paper presents the first study of edge transport barrier formation dynamics using a first-principles based electrostatic gyrokinetic simulation implemented in XGC1 in realistic edge geometry [30,31]. In the gyrokinetic equations, the fast gyro-motions are analytically treated, thus removing the gyrophase angle variable, while preserving the most basic plasma physics element at first principles level; i.e., the individual particle motions and their parallel Landau resonance with waves. Moreover, the XGC1 simulations evolve the total distribution function $f(\boldsymbol{x},\boldsymbol{v},t)$ for the gyrokinetic ions and the drift-kinetic electrons without scale separation, hence the background macro-scale kinetic neoclassical physics is self-consistently included together with the micro-scale nonlinear turbulence physics and no *apriori* linear instability drive assumption is made, except the low-beta electrostatic-limit assumption. In order to handle the orbit loss and non-Maxwellian physics properly, a conserving and fully nonlinear Fokker-Planck collision operator is used [32]. Lost plasma particles are recycled as Monte Carlo neutral atoms in the divertor chamber, with charge exchange and ionization interactions with plasma.

A global transport time-scale gyrokinetic investigation of the L-H transition (starting from a global L-mode transport equilibrium, gradually increasing the heating power to get the transition, and observing a pedestal build-up) is prohibitively expensive on the present-day leadership class computers.  In the present study, we make the simulation possible by reducing the computational resource requirement as much as possible via a model simplification; i.e., by choosing a fast electrostatic bifurcation case under strong forcing by a high rate of edge heat deposition without prolonging it to the slow, follow-on pedestal build up process. XGC1 simulations and analytic study show that edge turbulence saturation is usually established in ≲0.1ms [34,31], while in the core plasma nonlinear turbulence saturation is established in ≳1 ms.

By definition, a turbulence-bifurcating plasma is not in a global transport steady state. This implies that the establishment of a global transport steady-state may not be a necessary condition for a edge transport barrier formation study.  If the turbulence suppression in the edge layer can occur within <1 ms of plasma time by strong forcing, the transition dynamics can be studied on the 27 peta-flop-peak computer Titan at ORNL [33].

For the present study, we use the magnetic field geometry and the plasma profile from the Alcator C-Mod [35] L-mode plasma discharge #1140613017 as simulation inputs, but taking the toroidal magnetic field ($B_T$) to yield $V_B$ toward the magnetic X-point (i.e. the favorable direction for an H-mode transition); the actual discharge had $V_B$ away from the X-point. The plasma current is parallel to $B_T$. In these plasmas the electron kinetic energy density/magnetic energy density $\beta_e$ is only ≈ 0.01% just inside the separatrix and thus magnetic fluctuation effects are neglected.

We note here, however, that the electromagnetic effect may not be negligible in a real experimental L-H bifurcation dynamics.  How the magnetic fluctuations affect the present L-H bifurcation study is a subject of future study.

To minimize computational cost, an exaggerated amount of net heat $\Delta W_{layer}$≈0.8MW (significantly exceeding the experimentally observed net heat accumulation rate in the edge layer from the 1.6MW heat flux) is accumulated in the $0.947<\Psi_N< 0.989$ edge region so that the edge temperature is forced to increase at an exaggerated rate (Fig. 1), thereby quickly inducing edge transport bifurcation. $\Psi_N$ is a normalized minor radius in terms of poloidal magnetic flux that is zero on the magnetic axis and unity on the separatrix surface. The heat source is designed in the way not to generate an artificial flow in the plasma: After each heating time step in which a small fraction of the particle kinetic energy is increased, any momentum generation is removed by shifting back the particle distribution function in the parallel direction by a proper amount. Moreover, we applied the heat source only at $\Psi_N<0.76$ so that the heat accumulation in the edge region is from divergence of the radial heat flux. Note that the edge ion pressure gradient at and just inside $\Psi_N=1$ increases as the simulation proceeds. The edge electron temperature ($T_e$) also increases, and its gradient (not shown) actually steepens just inside $\Psi_N=1$.

A total-f gyrokinetic simulation always experiences oscillations in the transient geodesic acoustic modes (GAMs) [36-41] as the initial, approximate experimental plasma profile relaxes to a profile that is self-consistent with the gyrokinetic equilibrium and transport. These transient GAMs usually decay away after several oscillations in the near-equilibrium core plasma [36-41]. However, the GAM oscillation may persist longer or be easily excited in a transitional edge plasma due to weak poloidal winding of the edge magnetic field and a high free energy [31,42,43]. A strong GAM activity is indeed observed as the L-H bifurcation is approached in ASDEX-U [44].

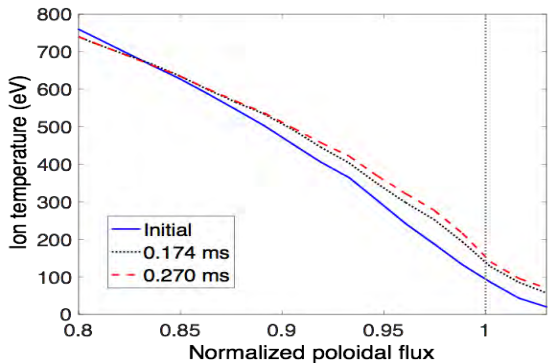

Fig. 1. Ion temperature ($T_i$) profile in the greater edge region of the modeled C-Mod plasma at three different times. Effect of the heating on $T_i$ can be seen to be significant in the edge layer.

Figure 2(a) depicts activities of the local ExB flow $V_E=-E_r/B$ (green dashed), its radial shearing rate $V_E' = dV_E/dr$ (red dotted), and the turbulence intensity $(\delta n/n)^2$ (blue) in the middle of the edge layer at $\Psi_N\approx0.975$. We will focus our attention to $V_E'$, not to $V_E$ itself, since the latter is found not directly correlated with the bifurcation event. Oscillations at approximately the theoretical GAM frequency of the modeled C-Mod edge plasma ($\tau_{GAM}\approx0.03$ms) can be observed in Fig. 2(a). Analysis of these oscillations show that they have an m/n=0/0 (velocity) and 0/1 (density) Fourier mode structure, consistent with GAMs. The electron heat flux is shown in Fig. 2(b) and also exhibits a similar oscillation which initially grows in amplitude. Figure 2(c) shows, together with Fig. 2(a), the $V_E'$ activities in radius-time with the initial transient shearing rate of ~$10^5$ Hz decreasing to a negligible level (~$10^4$ Hz) around 0.12ms as the edge turbulence is established, with its intensity modulated with the global GAM activity. We observe that the GAMs propagate from inner radii towards the edge, with a gradually decreasing radial propagation speed as they approach the edge and some interference pattern as they are reflected from the edge. A peculiar feature can be noticed in the $V_E'$ oscillation before t=0.175ms (denoted by vertical dash-dot line): Fig. 2(c) shows that in the edge layer near the magnetic separatrix, the positive peaks of the sheared ExB flow do not penetrate into the region $\Psi_N$>0.96, suggesting that at this time there is some mechanism at play to suppress the positive ExB shearing in this region.

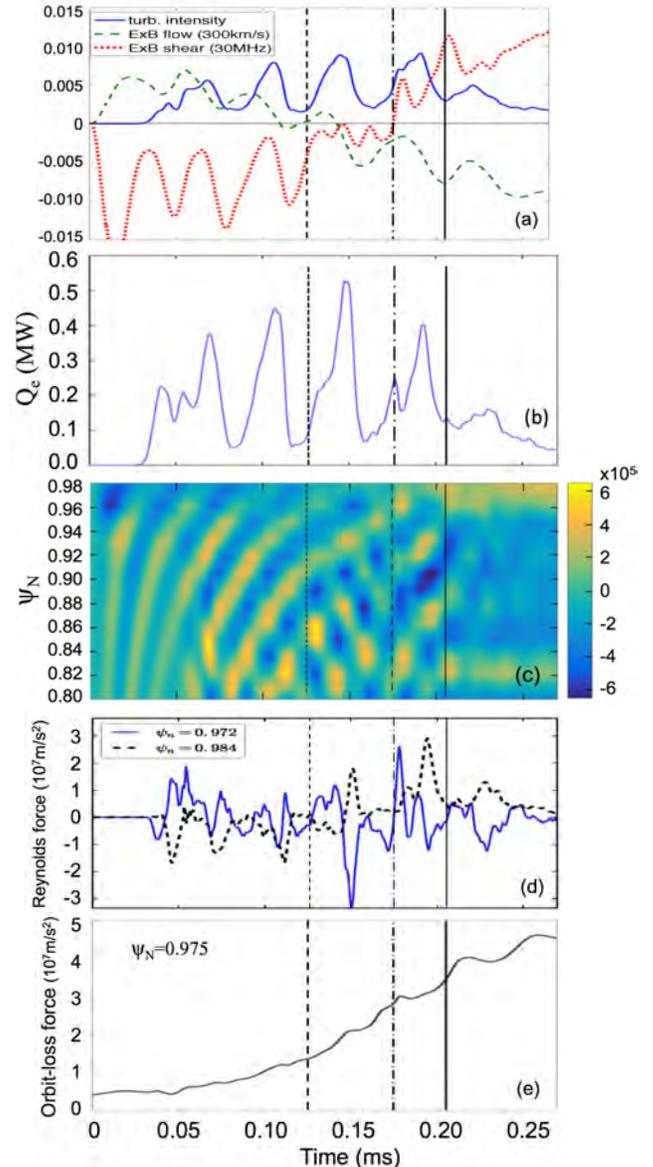

Fig. 2. Time behavior at $\Psi_N=0.975$ of (a) $(\delta n/n)^2$, ExB shearing rate, and ExB flow; (b) Electron heat flux; (c) ExB shearing rate in radius, (d) Reynolds force at $\Psi_N=0.972$ and $0.984$; and (e) orbit-loss force.

In the gyrokinetic Poisson equation [31] $(\rho_i^2/\lambda_{Di}^2)V_E' \approx n_e - n_{i,gc}$ in an L-mode edge, a negative ExB-flow shearing rate $V_E'<0$ implies that the guiding-center plasma is (slightly) positively charged in the edge layer 0.96<$\Psi_N$<0.98. This also implies that the electrons lead the particle loss, giving rise to a polarization response by ions.

Another critical feature can be seen in Fig 2. At $\Psi_N$=0.975, $V_E'$ oscillates while maintaining $V_E'<0$ prior to 0.12ms (vertical dashed line), and then between 0.12ms and 0.175ms as the nonlinear turbulence is established, $V_E'$~0 (Fig 2(a)). At 0.175ms the positive $V_E'$>0 oscillations begin to penetrate into the edge region $\Psi_N$>0.96 (Fig 2(c)), with $V_E'$ at $\Psi_N$≈0.975 increasing further in the positive direction (Fig 2(a)). $V_E'$ and $(\delta n/n)^2$ now show an out-of-phase, nonlinear limit cycle behavior. The peak shearing rate at $\Psi_N$≈0.975 exceeds ~300kHz at t≈0.205ms (solid vertical line), which coincides with the maximum linear growth rate of the most unstable dissipative modes [45] (i.e., dissipative trapped electron modes in the modeled plasma). Also, the second kick into the positive $V_E'$ direction that peaks at ≈0.205ms (Fig 2(a)) penetrates deeper toward the separatrix $\Psi_N$>0.97 (Fig. 2(c)). Around this time, the GAM oscillations at $\Psi_N$<0.95 are dying out: Thus for t~0.205ms the stronger penetration of the positive

$V_E'$ in the region $\Psi_N$>0.97 is not driven by a stronger GAM activities from the core region. It can also be seen that the sign of the average $V_E'$ inside the edge layer 0.95<$\Psi_N$<1 changes at t≈0.175ms, indicating that an electron-dominated particle loss has changed into an ion-dominated loss. Notice here also that the important ExB shearing actions are confined to a thin edge layer around 0.96≲ $\Psi_N$ ≲0.98.

After 0.205ms the $V_E'$ oscillations cease but $V_E'$ grows continuously in the positive direction, and the turbulence is continuously decaying after ~0.22ms. The radial electron thermal flux (Fig 2b) and ion thermal and particle fluxes (not shown) also then decay in the same fashion. At this stage, $V_E'$>0 becomes part of the background mean ExB flow shear with a net negative charge (ion-dominated loss). We identify this event as the final stage of turbulence and transport bifurcation after which the pedestal grows to H-mode condition.

Questions that arise at this point include: 1) what triggers the sudden penetration of the strong $V_E'$>0 part of the GAM oscillations into the edge layer at t≈0.175 and again at ≈0.19 ms, 2) why does $V_E'$ and its oscillations stay positive after the 0.175ms, and 3) what maintains the positive ExB flow as the turbulence is suppressed?

Figure 2(d), which shows the Reynolds force [7, 9], $F_{Re}$ = -d<$\tilde{v}_r\tilde{v}_\theta$>dr at $\Psi_N$=0.972 and 0.984, offers an answer to the first question: There are spatially localized oscillations of $F_{Re}$ into the positive poloidal direction (electron diamagnetic flow direction) in the edge layer at t≈0.175 and 0.190ms,

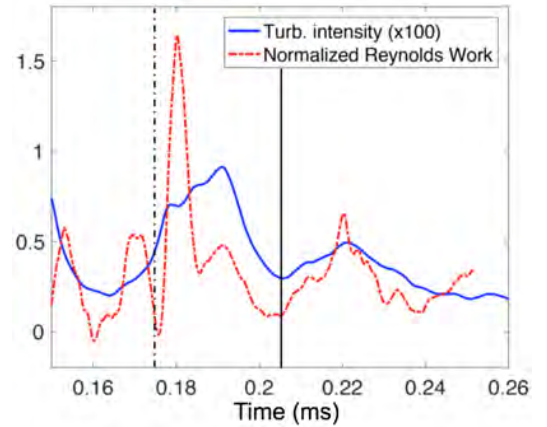

Fig. 3. Normalized consumption rate <$\tilde{v}_r\tilde{v}_\theta$>$V_\perp'$ /($\gamma_{eff}$ $\tilde{v}_\perp^2$/2) of the turbulence kinetic energy $\tilde{v}_\perp^2$/2 by the shearing $V_E'$ at $\Psi_N$=0.972 around the bifurcation time, using the critical shearing rate 300kHz as the effective source rate ($\gamma_{eff}$). Turbulence intensity dynamics is also shown.

with a radial gradient that promotes positive sheared flow $V_E'$ in the edge layer (since $dV_E'/dt \approx dF_{Re}/dr$).

The second and third questions imply that there is a background force, at this time, pushing the edge layer to a negative charge state or $V_E'>0$. The third question also suggests that this background force is strong enough to keep the turbulence suppressed in the edge layer.

The ion orbit loss mechanism in the presence of the magnetic X-point [28, 29] is a well-known and robust physics mechanism that drives the edge layer to a negative charge state or a $V_E'>0$ state. As the edge $T_i$ increases, the increasing ion orbit-loss phase-space hole provides a background force leading to a negative local charge with $V_E'>0$ that keeps the plasma losses ambipolar. The mechanism can also be interpreted as a loss-hole induced $J_r \times B$ return-current force on the main ions that drives a poloidal rotation profile until the viscous force balances the driving $J_r \times B$ return-current force in the H-mode equilibrium (see Fig. 8 of [28]).

Figure 2(e) shows a simple estimate of the underlying $J_r \times B$ return-current force, measured at $\Psi_N=0.975$, from the collisionless ion-loss hole in the vicinity of the magnetic X-point as function of time while the local $T_i$ increases from heating (Fig. 1). The orbit-loss driven $J_r \times B$ return-current force is comparable and adds to the Reynolds force of Fig. 2(d). The $V_E'$ behavior in Figs. 2 (a) and (c), and the second an third questions, can thus be understood as arising from the combined effects of the Reynolds force and orbit loss effects.

On these timescales, the diamagnetic component of $V_E'$ is still small compared to the total $V_E'>0$. A strong negative $V_E$ at $\Psi_N \approx 1$, or a negative $E_r$ well in the edge transition layer, has not formed yet either. With a weak $V_E \approx 0$ at $\Psi_N \approx 1$ and $V_E'>0$ in the edge layer, the edge electrostatic potential is in fact found to be positive around and right after the bifurcation time. As the edge pressure profile gradually steepens, a negative $E_r$ well will form and the usual H-mode pedestal structure is expected to emerge.

In order to test one of the most fundamental assumptions used in the predator-prey model [7, 9], the normalized consumption rate $P=<\tilde{v}_r \tilde{v}_\theta> V_E'/(\gamma_{eff} \tilde{v}_\perp^2/2)$ of the turbulence kinetic energy per unit mass, $\tilde{v}_\perp^2/2$, by the $V_E'$ shearing action (i.e. the rate of Reynolds work) at $\Psi_N=0.972$ is plotted in Fig. 3 around the bifurcation time, using the critical shearing rate 300kHz as the effective source rate $\gamma_{eff}$ of the turbulence kinetic energy. Turbulence intensity dynamics from Fig. 2(a) is also plotted for reference. It is indeed found that the rate of Reynolds work becomes momentarily large enough to consume a significant portion of the turbulence kinetic energy, as indicated by P>1 around t≈0.18ms and the cut-off of the top in the GAM-oscillating turbulence energy at the corresponding time. Moreover, the time integrated Reynolds work per unit mass after the transition (5.1 $m^2/s^2$) is somewhat greater than the maximal turbulence energy just before the transition (4.5 $m^2/s^2$).

In conclusion, a fast, forced bifurcation of turbulence and transport has been observed for the first time in an electrostatic nonlinear gyrokinetic simulation. The simulation shows validity of most of the underlying assumptions used by the popular predator-prey model, with one important addition that the neoclassical orbit loss physics also plays a critical role in the bifurcation process. We observe that an edge turbulence and transport bifurcation event occurs when the microscale turbulence-driven Reynolds force and the macroscale neoclassical orbit-loss force reinforce each other, and the combined ExB shearing rate in the edge layer reaches a critical level. Thus, the experimental argument based upon the orbit loss mechanism in [27] and the conventional Reynolds stress argument work together.

The present study indicates that an intrinsic limitation of the notion of Reynolds stress in the L-H bifurcation dynamics is its disappearance during the period of turbulence suppression, implying the necessity of some other mechanism for the generation of the sheared ExB flow to keep the turbulence suppressed while a high enough pressure pedestal is formed to provide the needed steady sheared ExB flow. Another limitation is in the lack of preferred direction in the Reynolds force (it fluctuates in both directions, see Fig. 2d). The synergistic orbit-loss driven ExB-shearing, caused by the rising edge $T_i$, that acts in the same direction as the steep $\nabla p_i$ driven ExB shearing that develops at a later time, provides such a mechanism, and may help reconcile some experimental observations that ascribe the transition to orbit loss effect [27] or neoclassical effect [46] with reports of the key role of turbulent stress [12-22]. There exist other experimental observations that identified a strong correlation between the L-H transition and the orbit loss driven ExB shearing rate [47, 48].

The spatial scale of orbit-loss physics is about the ion poloidal gyroradius ($\Delta\Psi \approx 0.05$), while that of the Reynolds stress variation is about $\Delta\Psi \approx 0.01$ (see Fig. 2d). The temporal scale of the orbit-loss force development is $\approx 0.05$ms and increasing (Fig. 2e), while that of the Reynolds stress is $\approx 0.01$ms (Fig. 2d) and fluctuating. Thus, the ion orbit-loss provides a background force, interacting with the space-time dynamical Reynolds force. The ion 90° collision time $\nu_i^c$ in the transition layer is $\approx 0.05$ms and similar to the ion orbit loss force time scale, and longer than the Reynolds stress time scale. The $\nu_i^c$ time could be related to the limit-cycle time scale (see Fig, 2a at t$\approx$0.17–0.21), but not conclusive due to the similarity with the GAM oacillation time-scale. The simulation time has to be longer than these time scales to study the L-H bifurcation dynamics (0.27ms here).

The synergism between the Reynolds and orbit loss forces is also consistent with the general experimental observation that the L-H bifurcation is more difficult in the case when the $\nabla B$-drift is away from the single-null magnetic X-point, in which the orbit-loss effect is weaker [28,29].

Authors thank L. Delgado-Aparicio for estimating the radiative power loss profile. This work is supported mostly by DOE FES and ASCR through the SciDAC program. Computing resources is provided by OLCF.

*Corresponding author: cschang@pppl.gov
†*Present Address: University of York, UK*